\documentclass[showpacs,twocolumn,pre]{revtex4}
\usepackage{psfrag,epsfig,amsfonts,amssymb,amsmath,graphicx,slashbox}
\usepackage{dcolumn}
\usepackage{dcolumn}

\newcommand{\steady}{\bar\rho }
\newcommand{\hr}{{\cal H}}
\newcommand{\ord}{{\cal O}}
\newcommand{\gset}{{\cal G}}
\newcommand{\tr}{\mbox{Tr}}
\newcommand{\da}{\Delta_{\! A}}
\newcommand{\dda}{\delta \! A}

\providecommand{\norm}[1]{\|#1\|}

\begin{document}

\title{Equilibration of Isolated Macroscopic Quantum Systems
under Experimentally Realistic Conditions}
\author{Peter Reimann}
\address{Universit\"at Bielefeld, Fakult\"at f\"ur Physik, 33615 Bielefeld, Germany}

\begin{abstract}
In how far does an non-equilibrium initial 
ensemble evolve towards a stationary long 
time behavior 
for an isolated macroscopic quantum system?
We demonstrate that deviations from a
steady state indeed become unmeasurably 
small or exceedingly rare after initial 
transients have died out under the 
following conditions:
The Hamilonian does not exhibit exceedingly 
large degeneracies of energy eigenvalues
and energy gaps.
A large number of energy levels 
is significantly populated by the
initial ensemble.
The system is observed by a measurement 
device with a reasonably bound working range 
compared to the resolution limit.
The entire experiment, ending with the
quantum mechanical measurement process,
can only be repeated a ``reasonable'' 
number of times.
It is argued that all these prerequisites
for equilibration are fulfilled under many, 
if not all, experimentally realistic 
conditions.
\end{abstract}

\maketitle

\section{Introduction}
\label{s1}
According to textbook Statistical Physics,
all properties of a macroscopic system at 
thermal equilibrium are perfectly described 
by the canonical 
ensemble when weakly coupled to a thermal 
bath and by the microcanonical ensemble 
when isolated from the rest of the world.
However, there still seems to be no truly 
satisfying explanation of why this 
is so \cite{skl93,pen79}.

Here we adopt the most common viewpoint
that investigations of this problem should 
be based on standard Quantum Mechanics 
and should start with the treatment of
isolated systems.
Indeed, it is widely believed that by 
considering a system in contact with a 
thermal bath as a single, isolated 
``supersystem'', the canonical 
formalism should be deducible from 
the microcanonical one.
In turn, to directly describe an open system 
alone (without the bath to which it is coupled)
by means of standard Quantum 
Mechanics does not seem possible.

Excellent reviews of the enormous number 
of pertinent ``classical'' (i.e. older) 
works are provided e.g. by \cite{skl93,pen79},
while more recent developments are 
covered e.g. by
\cite{spec,pol11}.
With our present work we summarize and
further develop a recently developed new 
approach 
\cite{rei08,lin09,lin10,rei10,sho11,sho12,rei12}
to the fundamental question:
In how far does an non-equilibrium initial 
ensemble evolve towards a stationary long 
time behavior?
Besides this problem of equilibration,
a second fundamental issue is the problem of 
thermalization, i.e.,
the question whether, and to what extend, 
such a steady state is
in agreement with the microcanonical ensemble
predicted by equilibrium Statistical Mechanics.
This important issue, either for an
isolated system {\em per se} or
for an isolated system-plus-bath 
composite, has been recently 
addressed e.g.\ in 
\cite{rei08,lin09,lin10,rei10,leb,%
tas98,sre99,caz06,rig07,kol07,man07,cra08,bar08,rig08,spec,%
rou10,pal10,bri10,gog11,ike11,ban11,pol11,cam11,jac11,%
kas11,ji11,can11,kas12,rig12}
but will not be considered here in any 
further detail.

A first key point of our approach is the 
modeling concept \cite{pen79} 
that any given experimental 
system gives rise to a specific, 
well defined statistical ensemble 
(density operator)
the details of which are, however, 
unknown in practice.
Typically, one only knows that 
in the initial state
certain (usually macroscopic) observables are
relatively sharply distributed about some
approximately known mean values.
A main challenge of the theory is to properly
cope with this lack of knowledge.
A second key point is not to modify or 
approximate in any way the exact Quantum Mechanical
time evolution.
The third key point is to focus on 
experimentally realistic observables,
exhibiting a finite range and a 
finite resolution.

For the rest, our present approach still 
covers essentially arbitrary (generic)
systems, and as such
is complementary to many recent 
investigations of different specific 
model classes, observables, and initial conditions
-- often with a main focus on the so-called eigenstate
thermalization hypothesis and
the role of (non-)integrability in this context
-- see e.g.\ 
\cite{tas98,sre99,caz06,rig07,kol07,man07,cra08,bar08,rig08,%
spec,rou10,pal10,bri10,gog11,ike11,ban11,pol11,cam11,jac11,%
kas11,ji11,can11,kas12,rig12}.
In particular, our present approach is not
restricted to macroscopic observables,
in contrast to e.g. \cite{khi60,wig71,leb}.
The reason is that equilibrium Statistical Mechanics 
in fact also covers microscopic observables, 
e.g. the position, velocity etc. of one 
specific ``tracer'' particle within an fluid, 
and its pertinent (probabilistic)
predictions are in perfect agreement with 
numerical simulations and with 
the rapidly increasing number of 
experiments on single molecules, 
nano-particles etc.

\section{Model assumptions}
\label{s2}
We consider a large (macroscopic but finite), 
isolated system, modeled 
in terms of a 
Hilbert space $\hr$
and a time-independent Hamiltonian $H :\hr\to\hr$ 
of the form
\begin{equation}
H = \sum_n E_n P_n \ ,
\label{1}
\end{equation}
where the $P_n$ are projectors onto the 
eigenspaces of $H$ with mutually different
eigenvalues $E_n$ and multiplicities
\begin{equation}
\mu_n:=\tr \{ P_n \} \ .
\label{5}
\end{equation}
For the sake of simplicity, we 
restrict ourselves throughout 
this paper to Hilbert spaces 
$\hr$ of arbitrarily large, but
finite dimensions.
Generalizations to infinite dimensions 
are readily possible along the lines of 
\cite{rei12}.
In particular, the number 
of distinct energy eigenvalues is given
by some finite integer $N_E$ and the 
indices $n$ in (\ref{1}) and (\ref{5})
run from $1$ to $N_E$.

Besides the multiplicities/degeneracies of the energy
levels from (\ref{5}), a related further quantity 
of central importance will be the maximal 
degeneracy of energy gaps
\begin{equation}
g := 
\max_{m\not = n}|\{(k,l) \, : \, E_k-E_l=E_m-E_n\}| \ ,
\label{67aa}
\end{equation}
where $|S|$ denotes the number of elements
contained in the (finite) set $S$.
In other words, $g$ is the maximal number of (exactly) 
coinciding energy differences among all possible pairs of 
distinct energy eigenvalues.
In particular, $g=1$ is tantanmount to the absence
of any degenerate energy gaps: For any
given $m\not=n$ (and thus $E_m\not=E_n$)
all energy gaps $E_k-E_l$ are different
from $E_m-E_n$, except for $k=m$ and $l=n$.

System states are described by
density operators $\rho (t)$, 
evolving according to
$\rho(t)=U_t\rho(0) U_t^\dagger$
with propagator
$U_t:=\exp\{-{\rm i} Ht\}$ and $\hbar =1$.
With (\ref{1}) we thus can conclude that
\begin{equation}
\rho(t)=\sum_{mn} \rho_{mn}(0) \exp[-{\rm i} (E_m-E_n)t] \ ,
\label{6}
\end{equation}
where we have introduced the auxiliary operators
\begin{equation}
\rho_{mn}(t):=P_m\rho(t)P_n \ .
\label{7}
\end{equation}

Observables are modeled by self-adjoint
operators $A$ with expectation values
$\tr\{\rho(t)A\}$.
Similarly as in (\ref{1}), any such $A$
can be written in the form
\begin{equation}
A=\sum_{\nu} a_{\nu}\, Q_{\nu} \ ,
\label{a1}
\end{equation}
where $Q_{\nu}$ are the projectors onto the eigenspaces of 
$A$ with $N_A$ mutually different eigenvalues $a_{\nu}$.

A single Quantum Mechanical measurement
process of the observable $A$ on the
system in the state $\rho(t)$ thus results
in one of the $N_A$ possible outcomes $a_{\nu}$, 
and the probability to obtain the specific 
outcome $a_{\nu}$ is given by
\begin{equation}
q_{\nu}(t):=\tr\{\rho(t)Q_{\nu}\} \ .
\label{a3}
\end{equation}
Upon repeating the same experiment
(same measurement and system state)
$N_{rep}$ times, every outcome $a_{\nu}$ is thus
obtained about $N_{\nu}:=q_{\nu}(t)N_{rep}$ times with
typical statistical fluctuations/uncertainties  
of the order of $\sqrt{N_{\nu}}$.
In other words, the ``true'' values of 
$q_{\nu}(t)$ can be determined by means of 
$N_{rep}$ measurements up to a statistical 
uncertainty of about
\begin{equation}
\delta q_{\nu}(t)=\sqrt{q_{\nu}(t)/N_{rep}} \ .
\label{a4}
\end{equation}

While in principle, one and the the same experiment 
can be repeated arbitrarily often, in practice
(under experimentally realistic conditions)
the number of repetitions $N_{rep}$
must remain ``reasonable''.
To be on the safe side, we henceforth take
for granted the quite weak bound
\begin{equation}
N_{rep}\leq 10^{20} \ .
\label{rep}
\end{equation}
In fact, also $N_{rep}\leq 10^{100}$ 
or $N_{rep}\leq 10^{1000}$ would still
be no problem in our later considerations,
and it seems convincing that a theory
which does not go very much beyond 
that will do.

As pointed out e.g. in
\cite{khi60,realobs2,realobs3,realobs4,realobs5,lof,geo95,pop06}
it is not necessary to admit any arbitrary self-adjoint 
operator $A$ in order to model real experimental 
measurements.
Rather, it is sufficient to focus on
experimentally realistic observables 
in the following sense \cite{rei08,rei08a,rei10}:
Any observable $A$ must represent an experimental 
device with a finite range of possible 
outcomes of a measurement,
\begin{equation}
\da := 
\max_{\norm{\psi}=1} \langle\psi|A|\psi\rangle
- \min_{\norm{\psi}=1} \langle\psi|A|\psi\rangle
= a_{max} - a_{min} \ ,
\label{range}
\end{equation}
where the maximization and minimization is over 
all normalized vectors $|\psi\rangle\in \hr$
and where $a_{max}$ and $a_{min}$ are the 
largest and smallest eigenvalues of $A$.
Moreover, this working range $\da$
of the device must be limited to experimentally 
reasonable values compared to its 
resolution limit $\dda$.
Due to the manifold and often not very well
understood contributions to the uncertainty
$\dda$ of a real experimental measurement, 
quantitative estimates are not easy.
Yet it seems reasonable to claim that 
uncertainties of less than, say $10^{-20}$,
of the instrumental range $\da$ are 
impossible to achieve, i.e.
\begin{equation}
\da/\dda
\leq 10^{20} \ .
\label{9'}
\end{equation}
Similarly as below (\ref{rep}), also
exponents of 100 or 1000 are still 
admissible, and to go even far beyond
that seems not necessary to
faithfully model any real or
numerical experiment.

A particularly interesting and important 
example is a digital instrument,
displaying $d_A$ decimal places.
In this case, we have that $N_A=10^{d_A}$ and 
we can assume without loss of generality that 
$a_{\nu} = \dda\,(\nu-1)$ with $\nu=1,...,N_A=10^{d_A}$
and thus $\da= (N_A-1)\,\dda$.
It is quite convincing that every experimentally 
realistic measurement device, both digital and
analoge, can be satisfactorily modeled along these lines.
Moreover, all measurements known to the present 
author yield less than $20$ relevant digits, i.e.\ 
\begin{equation}
\da/\dda = N_A -1 
\leq 10^{20} \ ,
\label{9''}
\end{equation}
in accordance with our general estimate (\ref{9'}).

Next we focus on the specific observable $A=P_n$,
describing according to (\ref{1})
the population of the (possibly degenerate)
energy level $E_n$ with expectation value 
(occupation probability)
\begin{equation}
p_n:=\tr\{ P_n \rho(t) \} \ .
\label{11}
\end{equation}
Note that $P_n$ commutes with $H$ from (\ref{1})
and hence the level populations $p_n$ are 
conserved quantities.

For a system with $f$ degrees of freedom
there are roughly $10^{\ord(f)}$ energy eigenstates 
with eigenvalues in every interval of $1$J beyond the ground 
state energy, see e.g.  \cite{lldiu} or section 2.1 in \cite{rei10}.
The same estimate carries over to the number
of energy eigenvalues under the assumption that
their multiplicities (\ref{5})
are much smaller than $10^{\ord(f)}$.
For a macroscopic system with
$f=\ord(10^{23})$, the energy levels are thus 
unimaginably dense on any decent energy 
scale and even the most careful experimentalist 
will not be able to populate only a few of them
with significant probabilities $p_n$,
in particular after averaging over 
the ensemble, i.e. over of many 
repetitions $N_{rep}$
of the experiment.
In the generic case we thus can conclude 
\cite{rei08,rei08a,rei10}
that even if the system's energy is fixed up 
to an extremely small experimental uncertainty,
and even if the energy levels are populated 
extremely unequally, we still expect that 
even the largest ensemble-averaged 
level population $p_n$ will be extremely 
small (compared to $\sum_n p_n=1$)
and typically satisfy the very
rough estimate
\begin{equation}
\max_n\, p_n=10^{-\ord(f)} \ .
\label{12}
\end{equation}

\section{Definition of equilibration}
\label{s3}
Generically, the statistical ensemble 
$\rho(t)$ is not stationary right from 
the beginning, in particular for an initial
condition $\rho(0)$ out of equilibrium.
But if the right hand side of (\ref{6})
depends on $t$ initially, it cannot 
approach for large $t$ any time-independent 
``equilibrium ensemble'' whatsoever.
In fact, any mixed state $\rho(t)$ 
returns arbitrarily close (with respect to some 
suitable distance measure in Hilbert space) 
to its initial state $\rho(0)$ for certain, 
sufficiently large times $t$, 
as demonstrated for instance in 
appendix D of reference \cite{hob71}.
We emphasize, that these arbitrarily close
recurrences do not refer to pure states 
only (as in the classical Poincar\'e recurrences)
but rather to arbitrary statistical 
ensembles $\rho(t)$.

As an example, we focus on the simplest case
that all eigenvalues $E_n$ in (\ref{1})
are non-degenerate and thus every projector
$P_n$ is of the form $|n\rangle\langle n|$,
where $|n\rangle$ is the (unique)
eigenvector belonging to the 
eigenvalue $E_n$.
Next, we consider any $\rho(t)$
which is (at least initially)
not completely independent of $t$.
So, according to (\ref{6}) there must 
exists a pair of indices
$m,\, n$ with the property that 
$\langle m|\rho(0)|n\rangle\not=0$
and $\Omega:=[E_n-E_m]/\hbar\not =0$.
Then, by focusing on the specific observable 
\begin{equation}
A=B + B^\dagger \ , \  B:=|m\rangle\langle n| /\langle m|\rho(0)|n\rangle
\label{29}
\end{equation}
it readily follows from (\ref{6}) that 
\begin{equation}
\tr\{\rho(t)A\}=2\, \cos(\Omega t) \ .
\label{30}
\end{equation}
In other words, the ensemble $\rho(t)$
exhibits permanent oscillations rather 
than equilibration, at least as far as 
the specific observable (\ref{29}) is 
concerned.

The main implication of the two previous
paragraphs is that equilibration 
cannot be true (and hence cannot be proven) 
in full generality and rigor.
Put differently,
Quantum Mechanics and equilibration
are strictly speaking incompatible.
Equilibration can at most approximately
hold true for a restricted class of 
observables $A$ and initial conditions 
$\rho(0)$.
We therefore focus on the
weaker notion of equilibration
from \cite{rei08,lin09,lin10,rei10,sho11,sho12,rei12},
which merely requires the
existence of a 
time-independent ``equilibrium state'' 
$\bar\rho $ (density operator) with
the property  that the difference
\begin{equation}
\sigma(t):=\tr\{\rho(t) A\}-\tr\{\steady A\}
\label{17}
\end{equation}
between the true expectation value
$\tr\{\rho(t) A\}$ and the 
equilibrium reference value
$\tr\{\steady A\}$ is unresolvably small
for the overwhelming majority 
of times $t$ contained in any sufficiently
large (but finite) time interval $[0,T]$.
In other words, the expectation values 
$\tr\{\rho(t) A\}$ may still exhibit 
everlasting small fluctuations around their 
``equilibrium values'' $\tr\{\steady A\}$, 
as well as very rare large excursions 
away from equilibrium 
(including the above mentioned recurrences \cite{pen79}), 
but quantitatively these 
fluctuations are either unobservably small
or exceedingly rare on any realistic time scale
after initial transients have died out.
(Note that those initial transients become 
irrelevant if $T$ is chosen large enough.)
The compatibility of the specific
example (\ref{29}), (\ref{30})
with this notion of equilibration will
be further discussed in Sect. \ref{s6}.

As we will see below,
such an equilibrium ensemble $\steady$ 
indeed exists under quite weak conditions, 
and is given by the following definition
\begin{equation}
\steady:=\sum_n\rho_{nn} \ ,
\label{18}
\end{equation}
where the operators $\rho_{nn}:=\rho_{nn}(t)$ are 
defined via (\ref{7}) and where
the time-arguments of $\rho_{nn}(t)$ 
have been omitted since these operators are in fact
time independent according to (\ref{6}), (\ref{7}).

Intuitively, (\ref{18}) may 
be viewed as the infinite-time average
of $\rho(t)$ from (\ref{6}), but also as 
the time-independent or the ``diagonal'' 
part of $\rho(t)$.
Yet it should be emphasized that the definition
(\ref{18}) is formally self-contained and 
unambiguous without any such ``interpretation''.
In particular, $\steady$ in (\ref{18}) is 
well-defined,
independently of whether the
infinite-time average of $\rho(t)$
actually exists or not.
Note that also physically, 
the ``equilibrium state'' 
$\bar \rho$ is mainly a theoretical ``vehicle'' 
without a direct correspondence in the real system.

In the following, we will see that the above
mentioned ``equilibration properties''
of $\sigma(t)$ from (\ref{17}) indeed 
can be demonstrated by adopting this specific 
choice (\ref{18}) of the ``equilibrium state'' 
$\bar\rho$.
To do so, the quantity of foremost 
interest turns out (as one might have 
expected) to be the time-averaged variance
\begin{equation}
\bigl\langle \sigma^2(t)\bigr\rangle_T 
:=\frac{1}{T}\int_0^T dt \, \sigma^2(t) 
\label{n1}
\end{equation}
following from (\ref{17}) and (\ref{18}).
Considering and estimating such averages
has a long tradition, see e.g. 
\cite{lud58,boc59,far64,per84,deu91,sre96,tas98}.
Substantial progress along this line
has been achieved quite recently in the works 
\cite{rei08,lin09,lin10,rei10,sho11,sho12,rei12}.
In the following we summarize and 
further develop this approach.

\section{Equilibration of mean values}
\label{s4}
Closely following the line of reasoning of
Short and Farrelly \cite{sho12}, 
we introduce (\ref{6}), (\ref{17}), (\ref{18}) 
into (\ref{n1}) to conclude that
\begin{equation}
\bigl\langle   \sigma^2(t)\bigr\rangle_T=
\Bigl\langle \Bigl|\sum_{m\not=n} \tr\{ \rho_{mn}  A\}
\exp\left[-{\rm i}(E_m-E_n)t\right]\Bigr|^2 \Bigr\rangle_T \ ,
\label{45}
\end{equation}
where $ \rho_{mn}(0)$ is abbreviated as
$ \rho_{mn}$ and the sum runs
over the finite set of pairs of labels 
\begin{equation}
\gset:=\bigl\{(m,n)\, : \, m ,n\in[1,\dots, N_E],\, m\not=n\bigr\} \ .
\label{46}
\end{equation}
Defining for any $\alpha\in\gset$ with $\alpha=(m,n)$ 
the two quantities
\begin{equation}
G_\alpha := E_m-E_n\ ,\qquad
v_\alpha := \tr\{\rho_{mn} A\} 
\label{48}
\end{equation}
and introducing the self-adjoint, non-negative
matrix $M$ with matrix elements
\begin{equation}
M_{\alpha\beta}:=\bigl\langle \exp\left[{\rm i}(G_\alpha-G_\beta)t\right]\bigr\rangle_T \,
\label{50}
\end{equation}
we can rewrite (\ref{45}) as
\begin{equation}
\bigl\langle \sigma^2(t)\bigr\rangle_T=
\Bigl\langle \Bigl|\sum_\alpha v_\alpha
\exp\left[-{\rm i} G_\alpha t\right]\Bigr|^2 \Bigr\rangle_T 
=\sum_{\alpha,\beta} v^\ast_\alpha M_{\alpha\beta} v_\beta \ .
\label{49}
\end{equation}
Denoting by $\norm{M}$ the standard operator norm of the 
matrix $M$ (tantamount to 
the largest eigenvalue of $M$)
it follows that \cite{sho12}
\begin{eqnarray}
\bigl\langle \sigma^2(t)\bigr\rangle_T \leq 
\norm{M}\, \sum_{\alpha} |v_\alpha|^2
= \norm{M}\, \sum_{m\not = n} \bigl|\tr\{\rho_{mn} A\}\bigr|^2 \ .
\label{51}
\end{eqnarray}

Obviously, every matrix element (\ref{50}) 
exhibits a well-defined limit as the averaging 
time $T$ (cf. (\ref{n1})) tends to infinity.
Thanks to our restriction to finite dimensions
below (\ref{1}) it then follows that also the 
norm $\norm{M}$ converges for $T\to\infty$.
Hence, $\norm{M}$ can be bounded from 
above for all sufficiently large 
(but finite) $T$.
Quantitatively, for non-degenerate energy gaps, i.e.
$g=1$ in (\ref{67aa}), one can infer from (\ref{50})
that $M$ approaches the unit matrix for $T\to\infty$
and hence $\norm{M}\to 1=g$.
As demonstrated in in \cite{sho12,rei12},
the same result $\norm{M}\to g$ for $T\to\infty$
remains valid for arbitrary $g$.
We thus can conclude that $\norm{M}\leq 2g$
for all sufficiently large $T$, 
where $g$ is the maximal degeneracy 
of energy levels from (\ref{67aa}).
Observing that the sum in (\ref{51})
can only grow by extending it to all 
$m ,n\in[1,\dots,  N_E]$ it follows that
\begin{equation}
\bigl\langle   \sigma^2(t)\bigr\rangle_T \leq
2\, g\,
\sum_{m,n} |\tr\{ \rho_{mn}  A\}|^2
\label{45c}
\end{equation}
for all sufficiently large $T$.
We remark that this is the only place where
our restriction to finite dimensions below 
(\ref{1}) is essential. While it is
quite plausible that everything still
``goes well'' in the limit of infinite 
dimensions \cite{rei08,rei10}, 
a more detailed justification is actually
not so obvious and has been 
worked out in \cite{rei12}.

Next we employ (\ref{7}) for 
$\rho_{mn}=\rho_{mn}(0)$, 
the property $P_m^2=P_m$
of any projector $P_m$, 
and the cyclic invariance of 
the trace to conclude that
\begin{eqnarray}
\tr\{ \rho_{mn}  A\} & = &
\tr\{  P_m  \rho(0)   P_n   A  P_m\} \ .
\label{53}
\end{eqnarray}
Since $\rho:=\rho(0)$ is a self-adjoint, 
non-negative operator,
one can infer that there
exists a self-adjoint, non-negative operator, which
we denote by $\rho^{1/2}$, and which satisfies the
relation $\rho^{1/2}\rho^{1/2}=\rho$.
Observing that
\begin{equation}
P_m  \rho   P_n   A  P_m 
= (  P_m  \rho^{1/2})( \rho^{1/2}   P_n   A  P_m)
\label{31}
\end{equation}
and exploiting the Cauchy-Schwarz inequality 
\begin{equation}
\bigl|\tr\{B^\dagger C\}\bigr|^2\leq \tr\{B^\dagger B \}  \tr\{C^\dagger C \}
\label{32}
\end{equation}
for the scalar product $\tr\{B^\dagger C\}$ 
of arbitrary operators $B$ and $C$
it follows with (\ref{7}), (\ref{11}), and (\ref{53}) that
\begin{equation}
|\tr\{ \rho_{mn}  A\}|^2
\leq   p_m 
\tr\{ \rho_{nn}   A   P_m  A\} 
\label{54}
\end{equation}
and hence
\begin{eqnarray}
\sum_{m,n} |\tr\{ \rho_{mn}  A\}|^2
& \leq &
\tr\{ (\sum_n \rho_{nn})   A  (\sum_m p_m  P_m)  A\} \ .
\end{eqnarray}
Introducing
\begin{eqnarray}
\omega & := & \sum_m p_m  P_m 
\label{45dz}
\end{eqnarray}
and taking into account (\ref{18}) we
thus obtain
\begin{eqnarray}
\sum_{m,n} |\tr\{ \rho_{mn}  A\}|^2
& \leq &
\tr\{ \bar\rho    A \omega  A\}
= \tr\{ \omega  A \bar\rho    A  \} \ .
\label{45d}
\end{eqnarray}
Finally, we observe that 
\begin{equation}
\tr\{B C\}\leq \norm{B}\, \tr\{C\}
\label{45d1}
\end{equation}
for arbitrary self-adjoint, non-negative operators
$B$ and $C$, where
\begin{equation}
\norm{B}:=\max\{ \langle\psi|B|\psi\rangle\, :\, \langle\psi|\psi\rangle=1\}
\label{45d2}
\end{equation}
denotes the standard operator 
norm (largest eigenvalue of $B$).
Since both $\omega$ and $A  \bar\rho   A$
are self-adjoint and non-negative,
it follows that
\begin{equation}
\tr\{ \omega  A \bar\rho    A  \} 
\leq
\norm{\omega}\,\tr\{A\bar\rho A\} 
=
\norm{\omega}\,\tr\{\bar\rho A^2\} \ .
\label{45d3}
\end{equation}
The first term on the right hand side
equals $\max_n p_n$ according to (\ref{45dz}).
In view of (\ref{45c}) and (\ref{45d}) 
we thus can conclude that
\begin{equation}
\bigl\langle   \sigma^2(t)\bigr\rangle_T
\leq  2\, 
\tr\{\bar\rho  A^2\} \,g\ {\max_n\,} p_n 
\label{45g}
\end{equation}
for all sufficiently large $T$.
Since both $\bar\rho $ and $A^2$ are self-adjoint
and non-negative we can exploit (\ref{45d1})
to conclude that 
$\tr\{\bar\rho  A^2\}\leq \norm{A^2}\,\tr\{\bar\rho \}$.
With $\tr\{\bar\rho \}=1$ it follows that
$\tr\{\bar\rho  A^2\}\leq \norm{A^2}$.
Next we note that $\sigma(t)$ in (\ref{17}) and thus
$\bigl\langle \sigma^2(t)\bigr\rangle_T$ in (\ref{n1})
remain unchanged if $A$ is replaced by $A+c\, 1$,
where $c$ is an arbitrary real number and $1$ the
identity operator.
Choosing $c$ so that $a_{min}=-a_{max}$ and thus 
$a_{max}=\da/2$ in (\ref{range}) it follows with
(\ref{45d2}) that $\norm{A}=\da/2$.
Altogether, we finally obtain the result 
\begin{equation}
\bigl\langle   \sigma^2(t)\bigr\rangle_T
\leq  \frac{1}{2}\, 
\da^2\, g\ {\max_n\,} p_n 
\label{45h}
\end{equation}
for all sufficiently large $T$.

Next we define for any given $\dda>0$ and $T>0$
the quantity
\begin{equation}
T_{\dda} := 
\big| \{0< t  <T\, : \, |\tr\{\rho(t)A\}-\tr\{\steady A\}|\geq\dda\}\big| \ ,
\label{69}
\end{equation}
where $|S|$ denotes the size (Lebesgue measure) of the set $S$.
According to (\ref{17}), $T_{\dda}$ thus
denotes the measure of all times $t\in [0, T]$ for which 
$|\sigma(t)|\geq \dda$ holds true.
It follows that $\sigma^2(t)\geq \dda^2$
for a set of times $t$ of measure $T_{\dda}$
and
$\sigma^2(t)\geq 0$ for all remaining times $t$
in $[0, T]$.
Hence the temporal average (\ref{n1}) must be at least
$\dda^2 T_{\dda}/T$ and 
we can conclude from (\ref{45h}) that
\begin{equation}
\frac{T_{\dda}}{T}\leq  \frac{1}{2}\, 
\,\left(\frac{\da}{\dda}\right)^2\, g\ 
{\max_n\,} p_n 
\label{71}
\end{equation}
for all sufficiently large $T$.
This is the first main result of our present
paper.
A more detailed discussion of its physical 
content is postponed to Section \ref{s6}.

\section{Equilibration of individual measurement outcomes}
\label{s5}
Next we focus on the special case that $A$ 
happens to coincide with one of the projectors 
$Q_{\nu}$ in (\ref{a1}). 
Similarly as in (\ref{69}),
we define for any given $\epsilon>0$ and $T>0$
the quantity
\begin{equation}
T_{\epsilon,\nu} := 
\big| \{0< t  <T\, : \, |\tr\{\rho(t)Q_{\nu}\}-\tr\{\steady Q_{\nu}\}|\geq\epsilon\}\big| \ ,
\label{69'}
\end{equation}
and similarly as in (\ref{71}) we can conclude 
from (\ref{45g}) and $Q_{\nu}^2=Q_{\nu}$ that
\begin{equation}
\frac{T_{\epsilon,\nu}}{T}\leq  \frac{2}{\epsilon^2}\, 
\tr\{\bar\rho Q_{\nu}\} \, g \ {\max_n\,} p_n 
\label{71'}
\end{equation}
for all sufficiently large $T$.

Returning to general observables $A$ of the form
(\ref{a1}) we denote, similarly as in (\ref{a3}),
by
\begin{equation}
\bar q_{\nu}:=\tr\{\bar\rho Q_{\nu}\}
\label{a3'}
\end{equation}
the probability that a single measurement of $A$
would result in the outcome $a_{\nu}$ if the 
system were in the state $\bar\rho$.
Similarly as in (\ref{a4}), 
upon repeating the same experiment 
$N_{rep}$ times, the ``true'' frequency $\bar q_{\nu}N_{rep}$ 
of the outcome $a_{\nu}$ can thus be determined up to a 
statistical uncertainty of about
\begin{equation}
\delta\bar q_{\nu}:=\sqrt{\bar q_{\nu}/N_{rep}} \ .
\label{a4'}
\end{equation}

Focusing on the special choice $\epsilon=\delta\bar q_{\nu}$ 
we can conclude from (\ref{71'})-(\ref{a4'}) that
\begin{equation}
\frac{T_{\delta\bar q_{\nu},\nu}}{T}
\leq  
2\, N_{rep} \, g \ {\max_n\,} p_n 
\label{72}
\end{equation}
for all sufficiently large $T$.
According to  (\ref{69'}), the left hand side
of (\ref{72}) quantifies the fraction of all times 
$t\in[0,T]$ for which one may find by means of $N_{rep}$ 
measurements a statistically significant difference 
between the two frequencies
$q_{\nu}(t)=\tr\{\rho(t)Q_{\nu}\}$ and 
$\bar q_{\nu}=\tr\{\bar\rho Q_{\nu}\}$
of observing the outcome $a_{\nu}$.
For all other times $t\in[0,T]$, 
the ``true'' frequency 
$q_{\nu}(t)$ cannot be distinguished from the
one theoretically predicted by means of 
$\bar\rho$.
Denoting by $T^{\ast}$ the measure of all
times $t\in[0,T]$ for which such a distinction 
is possible at least for one 
of the $N_A$ different indices $\nu$,
one readily sees that
$T^{\ast}\leq\sum_{\nu} T_{\delta\bar q_{\nu},\nu}$
and thus with (\ref{72}) that
\begin{equation}
\frac{T^{\ast}}{T}
\leq  
2\, N_{rep} \, N_A \, g\ {\max_n\,} p_n 
\label{73}
\end{equation}
for all sufficiently large $T$.
For all other time $t\in[0,T]$, it is 
impossible to observe by means 
$N_{rep}$ measurements of $A$
any kind of statistically significant 
difference between $\rho(t)$ 
and $\bar\rho$.

\section{Discussion}
\label{s6}
The first main result of our paper is 
the relation (\ref{71}). 
In fact, the same result has been obtained 
already before by Short and Farrelly in \cite{sho12}, 
however by means of a quite different line of 
reasoning\footnote{In fact, the estimate obtained in 
\cite{sho12} is even somewhat tighter than 
that in (\ref{71}).}.
In particular, these authors
focus in a first step solely on pure states
and only in the end extend their result 
to mixed states via purification.
On the other hand, our present line
of reasoning is very similar to the one 
from \cite{rei12}, but the final result 
is significantly stronger.

According to (\ref{69}), the left hand side 
of (\ref{71}) represents the fraction of all times
$t\in[0, T]$ for which there is an experimentally
resolvable difference between the true expectation 
value $\tr\{\rho(t)A\}$ and the time-independent 
equilibrium expectation value $\tr\{\steady A\}$.
On the right hand side of (\ref{71}), $\da/\dda$ 
is the range-to-resolution ratio,
which may be very large but still remains 
bounded according to (\ref{9'})
for experimentally realistic observables.
The next factor, $g$, is the maximal degeneracy 
of energy gaps from (\ref{67aa}).
Finally, ${\max_n\,} p_n$ represents the largest, 
ensemble-averaged occupation probability of 
the (possibly degenerate) energy 
eigenvalues $E_n$, see (\ref{11}).
Typically, one expects that the rough upper
bound (\ref{12}) applies, except if certain
energy eigenvalues
are so extremely highly degenerate that
the multiplicities defined in (\ref{5}) severely
reduce the pertinent energy level density
compared to the non-degenerate case, 
see above (\ref{12}).

For a system with sufficiently many degrees of 
freedom $f$ and no exceedingly large 
degeneracies of energy eigenvalues and energy 
gaps\footnote{Note that no such degeneracies 
are actually encountered in the generic case
\cite{per84,sre99,tas98,gol06,rei08,lin09,rei10},
i.e. in the absence of special reasons like
additional conserved 
quantities (e.g. due to (perfect) symmetries or 
when the system consists of non-interacting 
subsystems) or fine-tuning of parameters 
(``accidental degeneracies'').}, 
we thus can conclude from (\ref{71}) with 
(\ref{9'}) and (\ref{12}) 
that the system 
behaves with respect to 
any experimentally measurable expectation value
exactly as if it were in the equilibrium
state $\steady$ for the overwhelming majority of times
within any sufficiently large (but finite) 
time interval $[0, T]$.
In particular, $T$ must obviously be much larger
than the relaxation time in case
of a far-from-equilibrium initial 
condition $\rho(0)$.

Next we turn to the second main result
of our paper, namely Eq. (\ref{73}).
Its left hand side represents
the fraction of all times $t\in [0,T]$
for which one may find by means of $N_{rep}$
measurements of $A$ a statistically significant
difference between the ``true'' probabilities
$q_{\nu}(t)=\tr\{\rho(t)Q_{\nu}\}$ and
the corresponding ``equilibrium probabilities'' 
$\bar q_{\nu}=\tr\{\bar\rho Q_{\nu}\}$
of observing {\em any} of the
$N_A$ different possible outcomes $a_{\nu}$
of the Quantum Mechanical 
measurement process.
According to (\ref{rep}),
(\ref{9''}), and (\ref{12})
we can conclude from (\ref{73}) 
that for a system with sufficiently many degrees 
of freedom $f$ and no exceedingly large
degeneracies of energy eigenvalues and 
energy gaps, the true state $\rho(t)$
is practically indistinguishable from 
the equilibrium state $\bar\rho$ after
initial transients have died out.
More precisely, since a real experiment
can only be repeated a ``reasonable'' number 
of times (see (\ref{rep})),
the observable differences between
$\rho(t)$ and $\bar\rho$ are either 
smaller than the remnant statistical 
uncertainties or so unimaginably rare in 
time that they never actually occur 
in practice.

The main difference between (\ref{71})
and (\ref{73}) is that the former
refers to the expectation value of
$A$, i.e. the mean value over all
the actually observed outcomes
$a_{\nu}$ of the single repetitions of
the experiment.
In particular, statistical uncertainties 
due to finite numbers of experiments are
not taken into account, i.e. $N_{rep}\to\infty$
is implicitly taken for granted.
The result (\ref{73}) goes substantially
beyond (\ref{71}) in two respects.
First, it properly accounts for a
finite number of experiments.
Second and even more importantly,
not only the first moment (mean value) 
but actually the full probability distribution
of the different possible measurement outcomes 
is covered by the result (\ref{73}).

It is instructive to illustrate 
the above general considerations by means of
our specific example from (\ref{29}), (\ref{30}).
Recalling that in this example we are dealing with
a non-degenerate Hamiltonian (\ref{1})
with eigenvectors $|n\rangle$, 
one readily sees that the spectrum 
of $A$ from (\ref{29})
consists of the two (non-degenerate) 
eigenvalues  $a_{\pm}=\pm |\rho_{nm}(0)|^{-1}$,
and one (highly degenerate)
eigenvalue $a_0=0$.
In other words, we have $N_A=3$
distinct eigenvalues in (\ref{a1}).
Furthermore, one can infer, 
similarly as in (\ref{31}) and
exploiting (\ref{32}), that
$|\langle m|\rho(0)|n\rangle|^2
\leq \langle m|\rho(0)|m\rangle\, 
\langle n|\rho(0)|n\rangle$.
According to (\ref{11}), the latter product
equals $p_m\, p_n$.
It follows that $|a_{\pm}|\geq 1/\max_n p_n$ 
and hence with (\ref{range}) that
$\da\geq 2/\max_{n} p_{n}$.
For experimentally realistic initial 
conditions we can infer with (\ref{12})
that $\da\geq \ord (10^{f})$.
On the other hand, in order to 
resolve the oscillations from (\ref{30}),
the experimental uncertainty $\dda$ 
must not exceed unity.

In other words, while it is clearly impossible
to deny the existence of the permanently
oscillating expectation values (\ref{30}),
they cannot be actually ``seen'' without
admitting unrealistically large range-to-resolution 
ratios $\da/\dda$ or statistical ensembles with 
unrealistically large energy level populations 
$\max_n p_n$.

The same thing may alternatively also
be viewed as follows:
Any single (ideal) measurement process
always results
in one of the three outcomes
$a_+$, $a_-$, or $a_0$.
Recalling that  $a_{+}=-a_- \geq 1/\max_n p_n$
and $a_0=0$ it follows that
an infeasible number of measurements
$N_{rep}$ would be needed to resolve  
the order-one variations of the ensemble 
average (\ref{30}).

The above example also suggests that 
our assumptions of experimentally
realistic observables, number of measurements, 
and initial conditions
(or some similar restrictions)
are unavoidable for taming the 
oscillations in (\ref{6})
and thus overcoming the fundamental
incompatibility of basic Quantum Mechanics
with rigorous equilibration, as detailed 
in Sect. \ref{s3}.
In other words, it seems not an artifact of
our present (or of any other) specific
line of reasoning but rather a feature of 
``physical reality'' (within the realm of 
standard Quantum Mechanics) that invoking 
``our human inability'' is unavoidable to 
demonstrate and understand equilibration.

\subsection*{Acknowledgment}
Collaboration with Michael Kastner on
closely related issues is gratefully
acknowledged.
This work was supported by 
Deutsche Forschungsgemeinschaft 
under grant RE1344/7-1.

\end{document}